\documentclass[twocolumn,tighten]{aastex63}

\usepackage{graphicx}
\usepackage{xcolor}
\usepackage{booktabs}
\usepackage{amsmath}
\usepackage{color,soul}
\usepackage[flushleft]{threeparttable}

\shorttitle{Confirming discovery of CGM of NGC\,3221}
\shortauthors{Das et al.}
\graphicspath{{./}{figures/}}

\newcommand{\ovii}{\ion{O}{7}}
\newcommand{\oviii}{\ion{O}{8}}
\newcommand{\oviiin}{\ion{O}{8}}

\def\ovii{{{\rm O}\,{\sc vii}~}}
\def\oviii{{{\rm O}\,{\sc viii}~}}
\def\oviiin{{{\rm O}\,{\sc viii}}}

\def\suzaku{{\it Suzaku}}

\def\xmm{{\it XMM-Newton}}

\begin{document}

\title{The warm-hot, extended, massive circumgalactic medium of NGC\,3221: an \textit{XMM-Newton} discovery}

\correspondingauthor{Sanskriti Das}
\email{das.244@buckeyemail.osu.edu}

\author[0000-0002-9069-7061]{Sanskriti Das}
\affiliation{Department of Astronomy, The Ohio State University, 140 West 18th Avenue, Columbus, OH 43210, USA}
\author{Smita Mathur}
\affiliation{Department of Astronomy, The Ohio State University, 140 West 18th Avenue, Columbus, OH 43210, USA}
\affil{Center for Cosmology and Astroparticle Physics, 191 West Woodruff Avenue, Columbus, OH 43210, USA}
\author{Anjali Gupta}
\affil{Columbus State Community College, 550 E Spring St., Columbus, OH 43210, USA}
\affiliation{Department of Astronomy, The Ohio State University, 140 West 18th Avenue, Columbus, OH 43210, USA}

\begin{abstract}
\noindent 
Using \suzaku~data, we had found a $3.4\sigma$ evidence for the \textcolor{black}{X-ray emitting} warm-hot circumgalactic medium (CGM) in the L$^\star$ galaxy NGC\,3221. Here we present \xmm~data and outline an efficient, rigorous and well-defined method to extract the faint CGM signal. 
We confirm the CGM detection at $4\sigma$ significance within 30--200\,kpc of the galaxy. We claim with $99.62\%$ confidence that the CGM is extended beyond $150$\,kpc. The average temperature of the CGM is 2.0$^{+0.2}_{-0.3} \times 10^6$ K, but it is not isothermal. We find suggestive evidence for a  declining temperature gradient out to 125\,kpc and for super-virial temperature within 100\,kpc. While a super-virial temperature component has been detected in the Milky Way CGM, this is the first time a temperature gradient has been observed in the warm-hot CGM of any spiral galaxy. The emission measure profile is well-fit with either a $\beta-$ model or a constant density profile. Deeper data are required to constrain the temperature and density profiles. We also confirm the \suzaku~result that the warm-hot CGM is \textcolor{black}{one of the most massive baryon components} of NGC\,3221 and \textcolor{black}{can account for the missing galactic baryons}. 
\end{abstract}




\section{Introduction} \label{sec:intro}
\noindent It has been known from observations that the stellar and ISM (interstellar medium) components of nearby spiral galaxies account for a small fraction of their total baryons \citep{Tumlinson2017}, compared to the amount expected from the universal baryon fraction of $\Omega_b/\Omega_m = 0.157 \pm 0.001$ \citep{Planck2016}. A possible solution to this ``missing baryons" problem lies in the highly ionized warm-hot circumgalactic medium (CGM) extended out to the virial radius of a galaxy, as has been predicted by theoretical models \citep{White1978,Ford2014,Suresh2017}. \textcolor{black}{``Warm-hot" referrers to the temperature range of T$= 10^5$--$10^7$ K \citep{Cen1999}, the cooler part of which ($10^5$--$10^6$ K) is observable in the UV. The hotter part in the $10^6$--$10^7$ K range is believed to contain most of the baryonic mass \citep[e.g.,][]{Oppenheimer2016}. This} phase can be probed by highly ionized metals (e.g., \ovii and \oviiin), the dominant transitions of which lying in the soft X-ray band. \textcolor{black}{It is also imperative to define the term CGM. The CGM is generally considered to be the gaseous medium outside the disk of a spiral galaxy, extending out the virial radius, or even twice the virial radius \citep{Oppenheimer2018}. This region, however, is neither homogeneous nor isotropic. Strong nuclear outflows are detected out to several kpc from the disks of starburst galaxies like M82 and NGC\,1482 \citep[e.g.,][and references therein]{Strickland2004a,Strickland2004b}. Extra-planar diffuse emission is also detected around star-forming galaxies out to a few kpc from the disk midplane \citep{Strickland2004a,Tullmann2006,Yamasaki2009,Li2013,Bogdan2015,Hodges-Kluck2018}. These are important features of galaxies, but we do not refer to them as the CGM. We call the volume filling diffuse warm-hot gaseous medium on larger scales (beyond $\approx 10$kpc) as the CGM. 
}

The X-ray emission from the CGM around spiral galaxies is extremely faint, which makes their detection challenging. The warm-hot CGM around the Milky Way has been detected in both emission and absorption, and it may account for the missing mass \citep{Gupta2012,Gupta2014,Nicastro2016b,Gupta2017,Das2019c,Das2019a}. For external galaxies, however, observations become much harder. 
The extended CGM in X-ray emission has been securely detected only around massive galaxies ($M_\star>2\times10^{11} M_\odot$), and only out to a fraction of their virial radii, with mass insufficient to close their baryonic budget
\citep{Anderson2011,Dai2012,Bogdan2013a,Bogdan2013b,Anderson2016,Bogdan2017,Li2017,Li2018}.

X-ray emission from the CGM of Milky Way-type L* galaxies would be even fainter, and harder to detect. Indeed, no CGM emission is detected around any such galaxy, with the exception of NGC\,3221. We observed NGC\,3221 with \suzaku~for a total of 120\,ks exposure time; the good time interval (GTI), however, was only 41 ks. Using these \suzaku~data, \cite{Das2019b} found the evidence of \textcolor{black}{X-ray emitting}  warm-hot CGM from the region within 27--150 kpc at $3.4\sigma$ significance. The mass of the detected warm-hot CGM, the largest baryonic component of the galaxy system, could account for the missing galactic baryons. This is the first external L* galaxy with the evidence of an extended warm-hot CGM and baryon sufficiency. 

In this paper, we present 37 ks of archival \xmm~data of NGC\,3221. The exposure time of the \xmm~observation is comparable to the GTI of the \suzaku~observation. Because of the larger field-of-view (FoV) and larger effective area of \xmm~compared to \suzaku, we can probe out to larger radii from the galactic center, and at higher S/N using \xmm~data. \cite{Tullmann2006} have also presented the same \xmm~observation; their focus, however, was on the extraplanar emission within 20 kpc of NGC\,3221. Our work instead is on the extended CGM from 30 to 200 kpc from the center of NGC\,3221.  

 Our initial goal was to confirm our \suzaku~ detection of the CGM of NGC\,3221, which we do. However, the \xmm~ data have yielded exciting new results that we present in \ref{sec:modeling}. We find, for the first time, suggestive evidence for a temperature gradient in the CGM. 
 
 Our paper is structured as follows: we describe the data reduction and spectral analysis in section \ref{sec:redans}, report the results, compare with our previous \suzaku~analysis of the same galaxy, and discuss its implications in section \ref{sec:result}. Finally, we summarize the paper and outline some of the future plans in section \ref{sec:conclude}.

\section{Data reduction and analysis}\label{sec:redans}
\noindent Our goal is to extract and analyze the diffuse X-ray emission spectrum from the circumgalactic region of NGC\,3221, observed with \xmm~EPIC-pn. 
\subsection{Data reduction}\label{sec:reduction}
We reduce the 36.9\,ks archival data (ObsID: 0202730101) using XMM-Extended Source Analysis Software (ESAS)\footnote{\url{ftp://xmm.esac.esa.int/pub/xmm-esas/xmm-esas.pdf}}. 

\begin{figure}
    \centering
    \includegraphics[trim= 10 20 10 0, clip, scale=0.45]{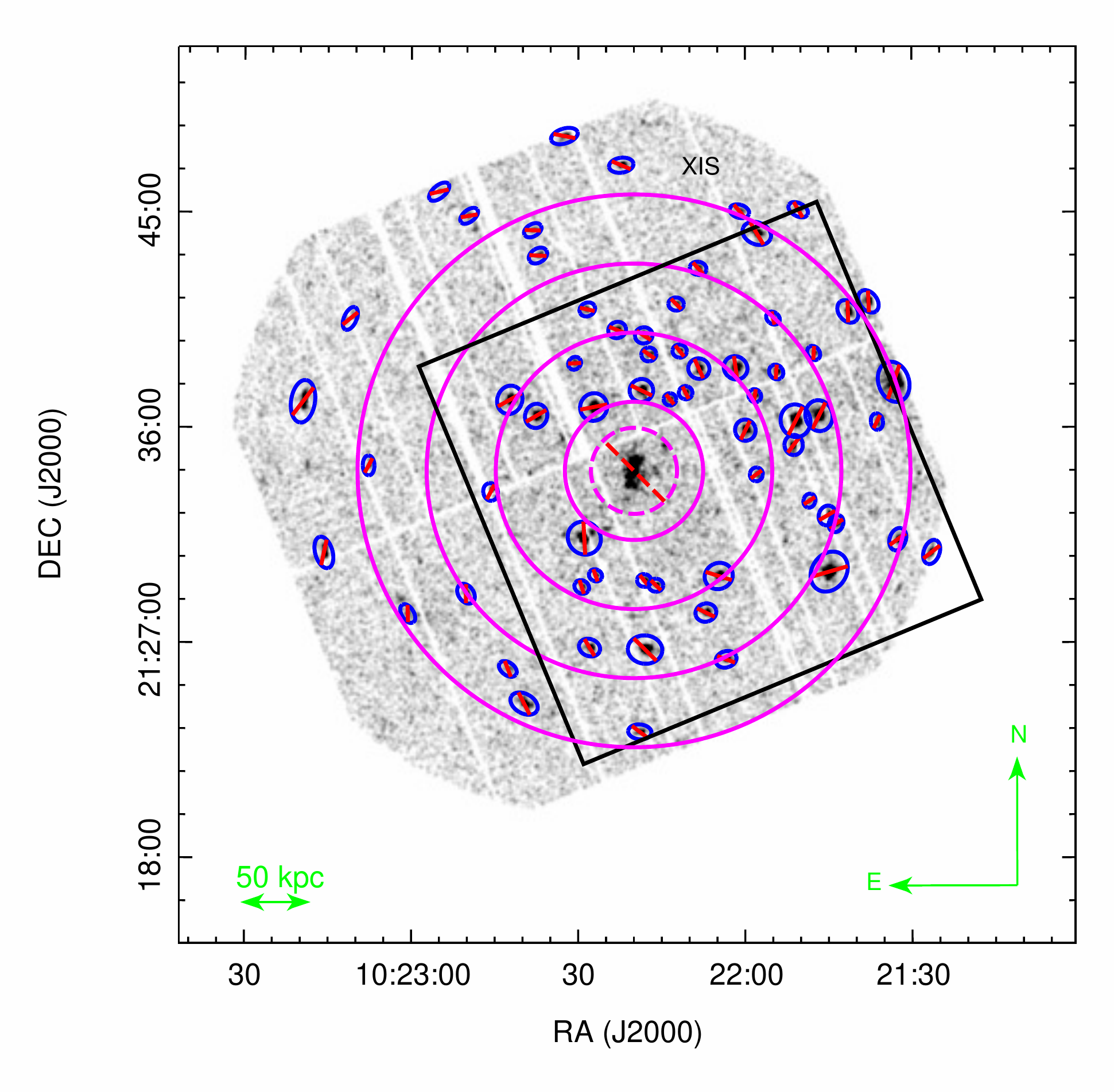}
    \caption{The image illustrating the galaxy field in 0.4--5.0\,keV. The detected and removed point sources are shown in blue circles. From inside to outside, the concentric magenta circles correspond to the 30, 50, 100, 150 and 200\,kpc radii around the disk of NGC\,3221. We consider the region beyond 200\,kpc to be the blank sky. The black tilted square marks the FoV of \suzaku/XIS.  The image is not background subtracted, and not exposure and vignetting corrected; the purpose of this figure is only to illustrate the regions of the annulii, the background, and the FOVs.  }
    \label{fig:image}
\end{figure}

We have worked with unfiltered event files and reprocessed data using the up-to-date calibration database. Below, we outline the data reduction procedure in brief: \\
1) We run \texttt{cifbuild} and \texttt{odfingest} to create the new CIF file and the observation summary file, respectively. \\
2) Filtering: For the pn data, we run the task \texttt{epchain} twice, one for the on-source and one for the OoT (out-of-time) exposure to process the event lists. Then, we run \texttt{pn-filter} to filter the data for soft proton (SP) flares, and create assorted diagnostic files. \texttt{pn-filter} calls the SAS task \texttt{espfilt}, which creates a high-energy (2.5-8.5 keV) count rate histogram from the FoV data, fits the peak with a Gaussian, determines thresholds at $\pm1.5\sigma$, creates a good-time-interval (GTI) file for those time intervals with count rates within the thresholds, and uses the task \texttt{evselect} to filter the data. We examine the count-rate histogram and the light curve of the FoV and the unexposed corners to make sure that there is no visually obvious residual contamination from the soft protons. The effective exposure time (GTI) after filtering is 24.7 ks, 67\% of the exposure time. It should be noted that \texttt{pn-filter} minimizes the SP contamination, but does not completely remove it. The residual SP contamination is modeled in the spectrum.\\
\textcolor{black}{MOS has better spectral resolution than pn, and it would have been nice to include MOS data in our spectral analysis. For the MOS data, we run \texttt{emchain} and \texttt{mos-filter}, which have the same function as \texttt{epchain} and \texttt{pn-filter} respectively. We found that the hardness ratio in 3 CCDs of MOS1 were unusually high, i.e., these CCDs were in anomalous state. For this reason we do not use MOS1 data in further analysis.} The effective area of MOS2 is smaller than pn by a factor of a few, therefore we analyze the pn data only to obtain larger photon count and hence, better signal-to-noise ratio (S/N). \\
3) Point source detection and excision: We detect the point sources using the routine \texttt{cheese}. As the effective area of pn falls off and the particle background is vary high beyond the 0.4-5.0 keV range, we detect the point sources and analyze the spectrum within this energy range.   We tune the following parameters in \texttt{cheese} to optimize the source detection and removal:\\ 
I) For GTI$\approx$ 25\,ks, Galactic N$_H = 1.76\times 10^{20}$ cm$^{-2}$ toward the direction of NGC\,3221 \citep{Bekhti2016} and an assumed $\alpha$= 1.7 powerlaw spectrum, the 4.5$\sigma$ sensitivity of pn is $\approx 10^{-16}$ ergs cm$^{-2}$ s$^{-1}$ \citep{Watson2001}\footnote{\textcolor{black}{The  appropriate limit for purely Poissonian background fluctuations to yield $\leqslant1$ spurious source per field is $\approx3.5-4\sigma$ \citep{Watson2001}. Therefore, 4.5$\sigma$ is a decent detectability criterion.}}. Therefore, the point-source flux threshold \textit{rate} is changed to 0.01 from the default value 1.0 (in the unit of 10$^{-14}$ ergs cm$^{-2}$ s$^{-1}$). This ensures that we identify sources down to 4.5$\sigma$ sensitivity limit. \\
II) The PSF threshold parameter \textit{scale} is changed to 0.15 from the default value 0.25. That means the point source is removed down to a level where the point source flux is 15\% of its maximum value instead of 25\%. This allows us to remove a larger fraction of the point source flux. Reducing the value further down to 0.10 did not produce any noticeable difference in the size and shape of the ellipses which mark the regions contaminated by point sources. \\
III) Minimum separation for point sources \textit{dist} is changed to 10$''$ from the default value 40$''$. This allows us to detect close-by sources. The on-axis PSF of Epic-pn is 12.5$''$ (FWHM), thus we ensure that all the resolved sources are counted. A value smaller  than 10$''$ did not detect any extra sources.  \texttt{cheese} takes into account the spatial variation of off-axis PSF over the detector plane; we do not assume or set any PSF value. 
\\
The data are carefully checked after source removal to make sure that any visibly identifiable source is not present. 
Additionally, we remove a circular region of $1.8'=30$\,kpc radius around NGC\,3221 (Figure \ref{fig:image}), which is slightly larger than the semi-major axis of NGC\,3221 in NIR \citep{Lehmer2010}.  This allows us to separate the large scale diffuse  X-ray  emission of the CGM from the emission from the stellar disk and the extra-planar region between 5.5 kpc (the semi-minor axis of NGC 3221) and 27 kpc (the semi-major axis of NGC\,3221).
\textcolor{black}{\subsection{Spectrum extraction}
\noindent Before going into detailed spectral analysis, we study the total intensity profile in the  0.4--5.0\,keV band. This includes the ``X-ray background" (XRB) in this FoV which we do not have any apriori knowledge about, and the warm-hot CGM, if any. 
\begin{figure}
    \centering
    \includegraphics[trim=20 5 40 40,clip, scale=0.475]{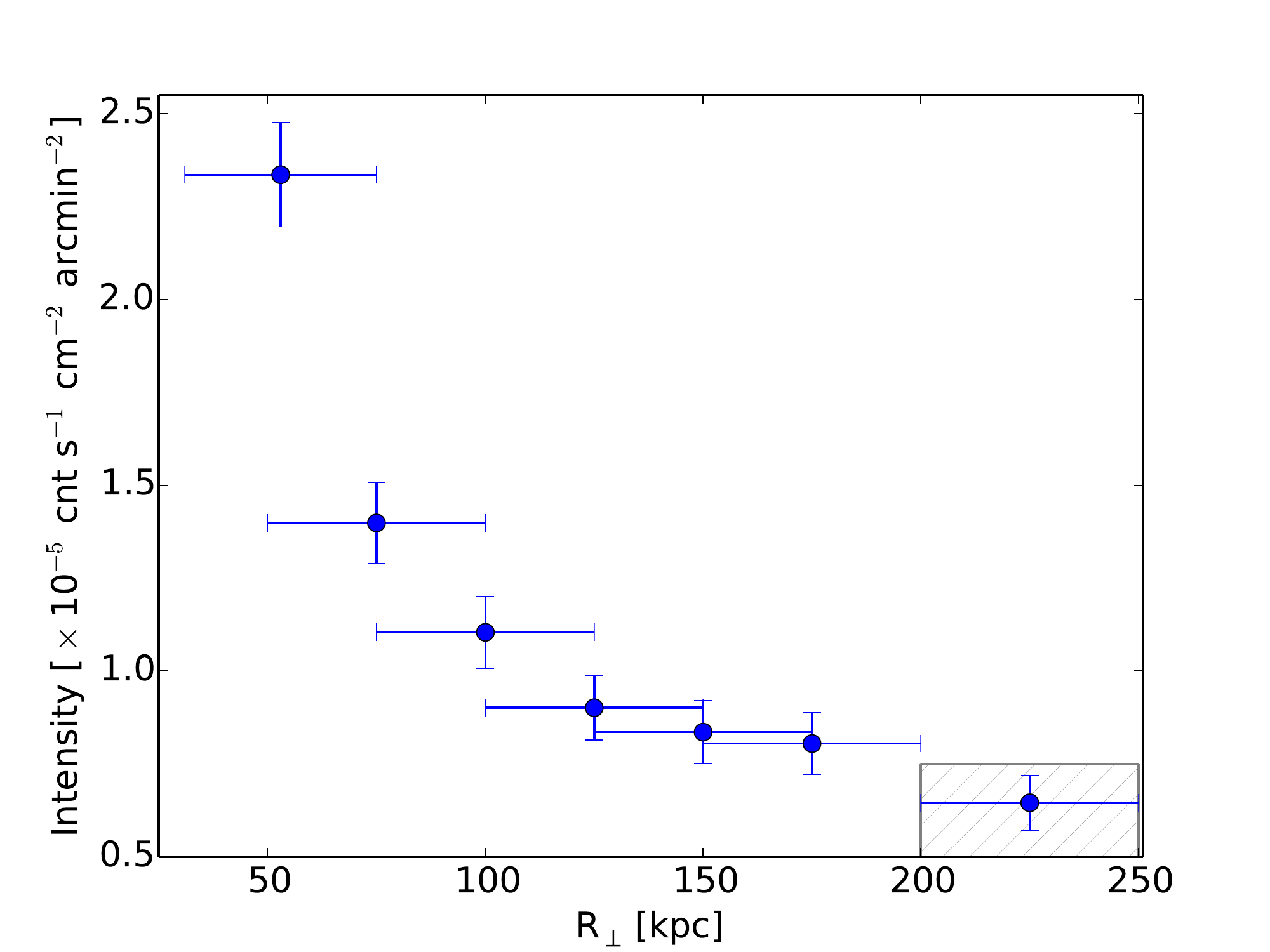}
    \caption{\textcolor{black}{Intensity profile in 0.4-5.0\,keV. The hatched region is used as ``\textit{blank sky}" in spectral analysis (\S\ref{sec:analysis}).}}
    \label{fig:Intensity}
\end{figure}
The intensity continues to fall with galactocentric radius (Figure \ref{fig:Intensity}). The XRB is very unlikely to show such a spatial correlation with the position of NGC\,3221. This indicates the presence of the CGM. 
}
We consider the region beyond 200 kpc from NGC\,3221 as ``\textit{blank sky}", and the region within 200 kpc from NGC\,3221 as ``\textit{on-source}". By definition this assumes that there is no detectable CGM emission beyond 200 kpc of the galaxy\footnote{If the CGM is truly extended out to the virial radius or beyond, the ``\textit{blank sky}" would have a CGM component in it. Therefore, our CGM detection is conservative; what we measure here is the difference in the CGM brightness between the halo within 200\,kpc and outside.}.

As noted above, the CGM emission is faint, therefore difficult to detect. For this reason, we have designed a novel approach to determine the confidence, overall detection significance and the spatial extent of the CGM. With this rigorous method, we could determine the temperature and the surface brightness profiles, even from shallow data.
We extract the ``\textit{on-source}" spectra from three sets of annuli: \\
1)\textbf{Seclusive}: annuli with outer radius (R$_{max}$) at 200\,kpc, and inner radii (R$_{min}$) at 50, 75, 100, 125 and 150\,kpc. To keep the solid angle and the collecting area (hence, photon count) as large as possible, R$_{max}$ is set at the largest value. If the CGM emission is detected beyond a given R$_{min}$ in an annulus, it assesses the \textbf{confidence} of the hypothesis that the CGM signal is present beyond that R$_{min}$. \\
2)\textbf{Cumulative}: annuli with R$_{min}$ at 30\,kpc and R$_{max}$ at 100, 125, 150, 175 and 200\,kpc. As the CGM is more likely to be detected closer to the disk than at large radii (also indicated by figure \ref{fig:Intensity}), R$_{min}$ is set at the smallest value possible. Larger R$_{max}$ would imply larger volume, and if the CGM emission signal is present, more photons from the CGM. If the detection significance gets saturated (or decreases) beyond a given R$_{max}$, it provides another measure of the spatial extent of the CGM. The annulus with the maximum detection significance would provide the \textbf{overall detection significance}.  \\ 
3)\textbf{Differential}: annuli of width 20--25\,kpc and 45--50\,kpc from the region 30--200\,kpc (see Table \ref{tab:table}). This provides us the radial profiles of temperature and surface brightness.

\textcolor{black}{In this way, we split the analysis in two parts: detection of the signal and physical characterization of the signal. After we are ``confident" about the ``significant" detection of the signal from ``seclusive" and ``cumulative" annuli, we extract the physical information from the ``differential" annuli. This makes the detection process more efficient, while the physical characterization using differential annuli is similar to previous studies.} 

We extract the spectrum, RMF and ARF using \texttt{pn-spectra} and extract the instrumental X-ray background using \texttt{pn-back} for each region mentioned above.
Each spectrum is binned using ftool \textit{grppha} such that minimum count in each bin is 50, which gives a moderate S/N.

\begin{figure*}
    \centering
    \includegraphics[trim=20 5 40 40,clip, scale=0.475]{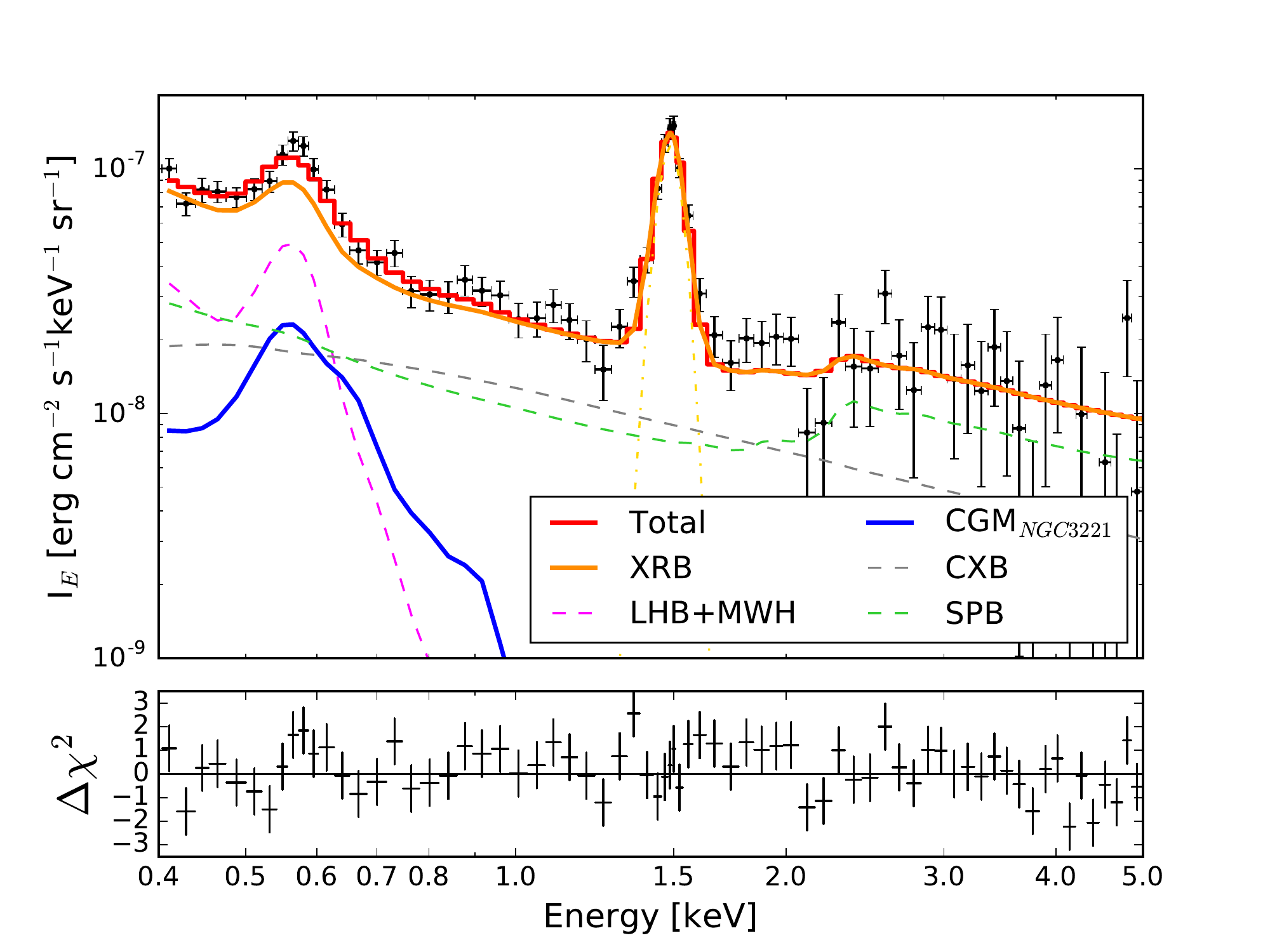}
    \includegraphics[trim=20 5 40 40,clip, scale=0.475]{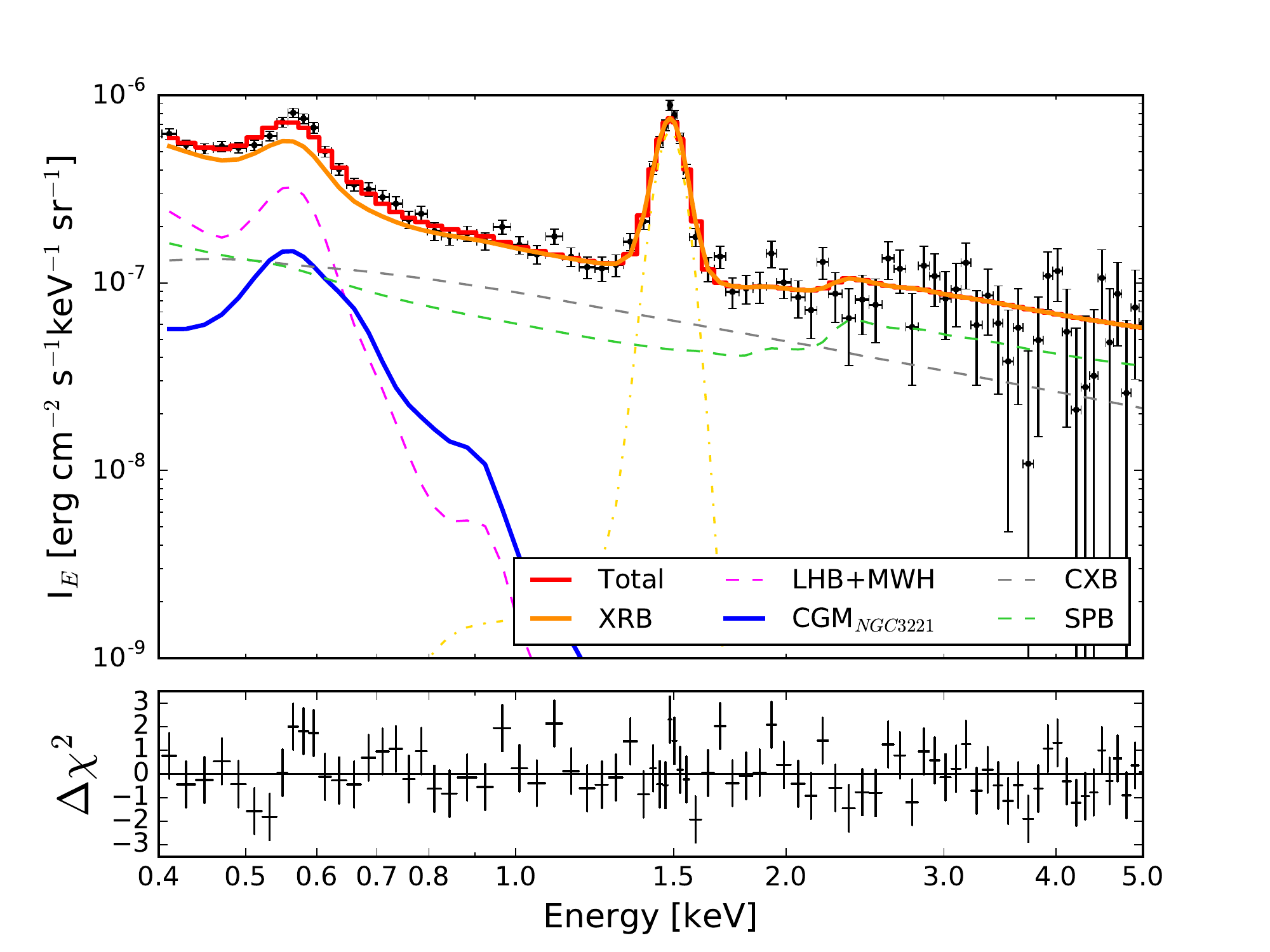}
    \caption{\textcolor{black}{Left: the spectrum from 150--200\,kpc region, right: the spectrum from 30--200\,kpc.} The best fit is shown by the red line and the CGM emission is shown by the blue line. The orange line corresponds to the X-ray background (XRB), i.e., the combination of the instrumental line, SPB, LHB, CXB and MWH. The offset between the red and the orange curves around 0.55--0.65\,keV is due to the \ovii and \oviii emission from the CGM of NGC\,3221.}
    \label{fig:spec}
\end{figure*}
\subsection{Analysis}\label{sec:analysis}
\noindent The ``\textit{blank sky}" spectrum contains both the foreground and background components. 
We obtain the best fitted spectral model for the ``\textit{blank sky}", and use that as the initial estimates of the ``X-ray background" (XRB) while analyzing the ``\textit{on-source}" spectra. All spectral analyses have been performed in \texttt{XSPEC}.

We model the ``\textit{blank sky}" spectrum as a combination of 4 components: 1) Instrumental Al K$\alpha$ line and soft proton background (SPB), 2) Local Hot Bubble (LHB), 3) Cosmic X-ray background (CXB) and 4) Galactic Halo, i.e., the warm-hot Halo of the Milky Way (MWH). This is similar to the model\,A of \cite{Das2019c}. Note that this X-ray ``background" is made of both foreground and background. As we are interested in any excess emission from the ``\textit{on-source}" spectrum compared to the ``\textit{blank sky}", the details of the XRB model should not matter as long as the same model is used for both the ``\textit{blank sky}" and the ``\textit{on-source}" spectrum (see appendix\,A of \cite{Das2019c} for details of the spectral model). 

The ``\textit{on-source}" spectrum consists of five components. The first four components are the same as those of the ``\textit{blank sky}" spectrum.  The fifth component is the CGM emission of NGC\,3221; we model it as a colliosionally ionized plasma in thermal equilibrium ({\bf apec} in XSPEC) at the  redshift of NGC\,3221, absorbed by the Galactic ISM. The free parameters of the model are the temperature and metallicity-weighted normalization factor. We
freeze the metallicity at solar. The normalization factor is \textcolor{black}{anticorrelated with metallicity}, so
the exact value of the input metallicity does not matter as long as the scaling is taken into account while quoting the metallicity-weighted values. (see \S3.2 of \cite{Das2019b} for details). 

We fit each ``\textit{on-source}" spectrum simultaneously with the ``\textit{blank sky}". The intensity of the Al K$\alpha$ line can change at different parts of the detector plane, so we do not tie its amplitude in the ``\textit{on-source}" and the ``\textit{blank sky}" spectra. We assume that the components of XRB (LHB, CXB, MWH, SPB) of the ``\textit{on-source}" spectrum are quantitatively the same as those of the ``\textit{blank sky}", with the normalization parameters scaled according to the ratio of projected areas of the ``\textit{on-source}" and the ``\textit{blank sky}" regions. As all spectra are extracted from the same observation and the same field, any spatial or temporal variations of the XRB are highly unlikely, thus validating our assumption.  

As the purpose of fitting the ``seclusive" and the ``cumulative"  annuli is to detect the signal from the CGM, we take an agnostic approach while modeling their XRB. For each annulus, we allow the XRB to vary, but tie it between the ``\textit{on-source}" and the ``\textit{blank sky}". Once the presence of the signal is confirmed, the XRB parameters obtained by fitting the largest annulus (30--200 kpc) are considered as the best estimate. We freeze the XRB parameters at these values while fitting the differential annuli. The best-fit spectrum in the 30--200 kpc region is shown in Figure \ref{fig:spec}.


We quantitatively measure the importance of the detection of the CGM emission from NGC\,3221 in two different ways, by measuring the \textbf{``confidence"} and the \textbf{``significance"}. If excess emission from the ``\textit{on-source}" spectrum is required in the spectral model, we call it ``confidence". The confidence is measured by performing the F-test (\texttt{XSPEC} command \texttt{ftest}). The difference in the $(\chi^2,dof)$ of the best-fitted models with/without the CGM component provides the F-statistic value and the null hypothesis probability P$_{null}$. The confidence of the presence of the CGM, as we discuss in the following section, is (1-P$_{null}$)$\times100$\%. The ``seclusive" annuli are used to confirm the presence of the signal. So for them, the relevant parameter is ``confidence".  The ``significance", on the other hand, refers to the statistical significance of a measured parameter. \textcolor{black}{We note the ``significance" of a detection only when we are ``confident" that the signal is required in the spectral model (Table \ref{tab:table}).} The ``cumulative" annuli are used to detect the signal. So for them, the relevant parameter is ``significance". We calculate the uncertainties of the temperature and the normalization of the CGM component using the \texttt{XSPEC} command \texttt{err} and \texttt{steppar} (when $\chi^2$ is not monotonic)\footnote{The values quoted in table \ref{tab:table} are the combination of systematic and statistical uncertainty. Systematic uncertainties are due to the XRB components, and it dominates the total uncertainty. The statistical uncertainty is larger in smaller annuli due to smaller photon count. The systematic uncertainty is larger in larger annuli because of possible variation of the XRB over the FoV. As the detection significance involves the total (systematic and statistical) uncertainty, we do not report the decomposed uncertainties to avoid complication.}. We define the detection significance as the ratio of the best-fitted value of the normalization parameter and its $1\sigma$ error in the lower end. 
Unless explicitly mentioned otherwise, we quote uncertainties as 1$\sigma$ error bars, and ranges as 68\% confidence intervals. 

\begin{table*}[ht]
\centering
\caption{Best fitted parameter values}
\label{tab:table}
\resizebox{\textwidth}{!}{%
\begin{tabular}{cccccc}
\toprule
\multicolumn{6}{c}{``seclusive" annuli*}\\
R$_{min}$ & R$_{max}$ & Norm &  \textcolor{black}{kT}    & Confidence      & \textcolor{black}{Significance}               \\
{[}kpc{]} & {[}kpc{]} & {[}$\times$10$^{-5}$cm$^{-5}${]}     & [keV]                 & {[}\%{]}  & [S/N]                       \\
\hline
50        & 200       & 8.6$^{+3.8}_{-2.2}$    & \textcolor{black}{$0.17^{+0.02}_{-0.03}$}            & \textbf{99.97}       & \textcolor{black}{3.9}                    \\
75        &           & 8.1$^{+4.1}_{-2.1}$  & \textcolor{black}{$0.17^{+0.02}_{-0.03}$}               & \textbf{99.96}       & \textcolor{black}{3.9}                       \\
100       &           & 7.1$^{+3.8}_{-1.8}$ & \textcolor{black}{$0.17^{+0.02}_{-0.03}$ }               & \textbf{99.94}       &    \textcolor{black}{3.9}                   \\
125       &           & 5.6$^{+3.1}_{-1.5}$  & \textcolor{black}{$0.17^{+0.02}_{-0.03}$ }              & \textbf{99.86}        & \textcolor{black}{3.7}                   \\
150       &           & 4.6$^{+9.0}_{-1.9}$   & \textcolor{black}{$0.14^{+0.03}_{-0.04}$ }             & \textbf{99.62}             & \textcolor{black}{2.4}              \\
\hline
\multicolumn{6}{c}{``cumulative"  annuli*}\\
R$_{min}$ & R$_{max}$ & Norm  &  \textcolor{black}{kT}    & \textcolor{black}{Confidence}    & Significance$\dagger$           \\
{[}kpc{]} & {[}kpc{]} & {[}$\times$10$^{-5}$cm$^{-5}${]}   & [keV]                 & {[}\%{]}                   & {[}S/N{]}                       \\
\hline
30        & 100       & 1.8$\pm0.7$   & \textcolor{black}{$0.20^{+0.03}_{-0.02}$}                     & \textcolor{black}{98.11}                   & \textbf{2.6 (2.6)}                       \\
          & 125       & 2.8$\pm1.0$    & \textcolor{black}{$0.19\pm0.03$}                          & \textcolor{black}{98.82}             & \textbf{2.8 (2.8)}                       \\
          & 150       & 4.5$^{+1.3}_{-1.4}$ & \textcolor{black}{$0.18^{+0.02}_{-0.03}$} & \textcolor{black}{99.66}                                  & \textbf{3.1 (3.4)}                       \\
          & 175       & 7.4$^{+3.7}_{-2.0}$ & \textcolor{black}{$0.17^{+0.02}_{-0.03}$ }      & \textcolor{black}{99.95}                           & \textbf{3.8 (4.4)}                      \\
          & 200       & 8.9$^{+3.7}_{-2.2}$ & \textcolor{black}{$0.18^{+0.02}_{-0.03}$}          & \textcolor{black}{99.95}                         & \textbf{4.0 (4.1)}                       \\
\bottomrule
\end{tabular}
\begin{tabular}{cccc}
\toprule
\multicolumn{4}{c}{Differential annuli*}\\
R$_{min}$ & R$_{max}$ & Norm                                                  & kT                              \\
{[}kpc{]} & {[}kpc{]} & {[}$\times$10$^{-6}$cm$^{-5}${]}                      & {[}keV{]}                       \\
\hline
30        & 50        & {1.9$^{+1.6}_{-1.1}$}                   & 0.20$^{+0.08}_{-0.07}$ \\
50        & 75        & {3.9$^{+1.4}_{-1.3}$}                   & 0.29$^{+0.10}_{-0.06}$ \\
75        & 100       & {27.5$^{8.1}_{-7.4}$}                  & 0.15$^{+0.02}_{-0.04}$  \\
30        & 100       & 33.3$^{+8.3}_{-7.6}\bullet$        & 0.22$^{+0.07}_{-0.05}$   \\
50        & 100       & 31.4$^{+8.2}_{-7.5}\bullet$         & 0.21$\pm0.03$ \\
30        & 75        & 6.1$\pm1.7$                     & {0.23$^{+0.07}_{-0.02}$} \\
75        & 125       & 23.4$^{+18.2}_{-7.4}$                & 0.16$^{+0.02}_{-0.03}$    \\
100       & 150       & 35.6$^{+109.5}_{-17.3}$               & 0.13$\pm0.04$ \\
125       & 175       & 40.9$^{+43.9}_{-15.1}$                & 0.14$\pm0.03$ \\
150       & 200       & 41.8$\pm7.1$                & 0.14$^{+0.03}_{-0.04}$ \\
125       & 200       & 50.6$^{+19.7}_{-7.7}$       & 0.17$^{+0.01}_{-0.02}$ \\
\bottomrule
\end{tabular}%
}
\footnotesize{*See \S\ref{sec:reduction} for definition}\\
$\dagger$Values quoted in parentheses are obtained by freezing the temperature at its best-fitted value from the region 30--200\,kpc. \\
$\bullet$ Obtained from the area-weighted average of the emission measures (i.e. the mean of the emission integrals) at 30--50, 50--75 and 75--100\,kpc. 
\end{table*}

\section{Results and discussion}\label{sec:result}
\subsection{Detection of the CGM emission}\label{sec:detection}
\noindent Using the extensive spectral analysis described above, we extract the signal from the CGM of NGC\,3221 with $>99.6\%$ confidence from the $150-200$\,kpc region (Table \ref{tab:table}, first segment). This confirms that the spatial extent of the CGM is $>150$\,kpc. The confidence increases as R$_{min}$ gets smaller, indicating the presence of the signal throughout the volume. 


Once the presence of the signal is confirmed, we determine the detection significance from the ``cumulative"  annuli. The significance continues to increase out to 200\,kpc (Table \ref{tab:table}, second segment), indicating that the CGM is likely extended out to 200\,kpc. We detect a $4\sigma$ signal from the 30--200\,kpc region (Figure \ref{fig:spec}). This confirms the detection by \cite{Das2019b}, but now with a higher significance. As the temperature and the normalization are strongly anti-correlated in our fits, we calculate the marginal detection significance by freezing the temperature at its best-fit value. This escalates the detection significance to 4.4$\sigma$ from 30--175\,kpc, and a 4.1$\sigma$ signal from 30--200\,kpc (Table \ref{tab:table}). This indicates that the CGM may be extended to 175\,kpc instead of 200\,kpc. However, given the small photon count, we do not fine-tune the spatial extent of the CGM of NGC\,3221; it is clearly extended beyond 150 kpc.   

\subsection{Confirmation as the CGM signal}\label{sec:confirm}

\noindent The statistical detection of the signal does not physically confirm it as the CGM of NGC\,3221, because the CCDs of EPIC-pn cannot really distinguish between the redshift $0$ and the redshift of NGC\,3221. However, we extract the XRB from the same field as the signal. The angular separation between the inner annulus and the ``\textit{``\textit{blank sky}"}" corresponds to a maximum physical length of 1\,kpc within the MWH. Therefore, if the detected signal is Galactic, it has to have a size $<$1\,kpc, which is very unlikely for the diffuse medium. 

\begin{figure}[h]
    \centering
    \includegraphics[trim = 10 0 0 30,clip,scale=0.45]{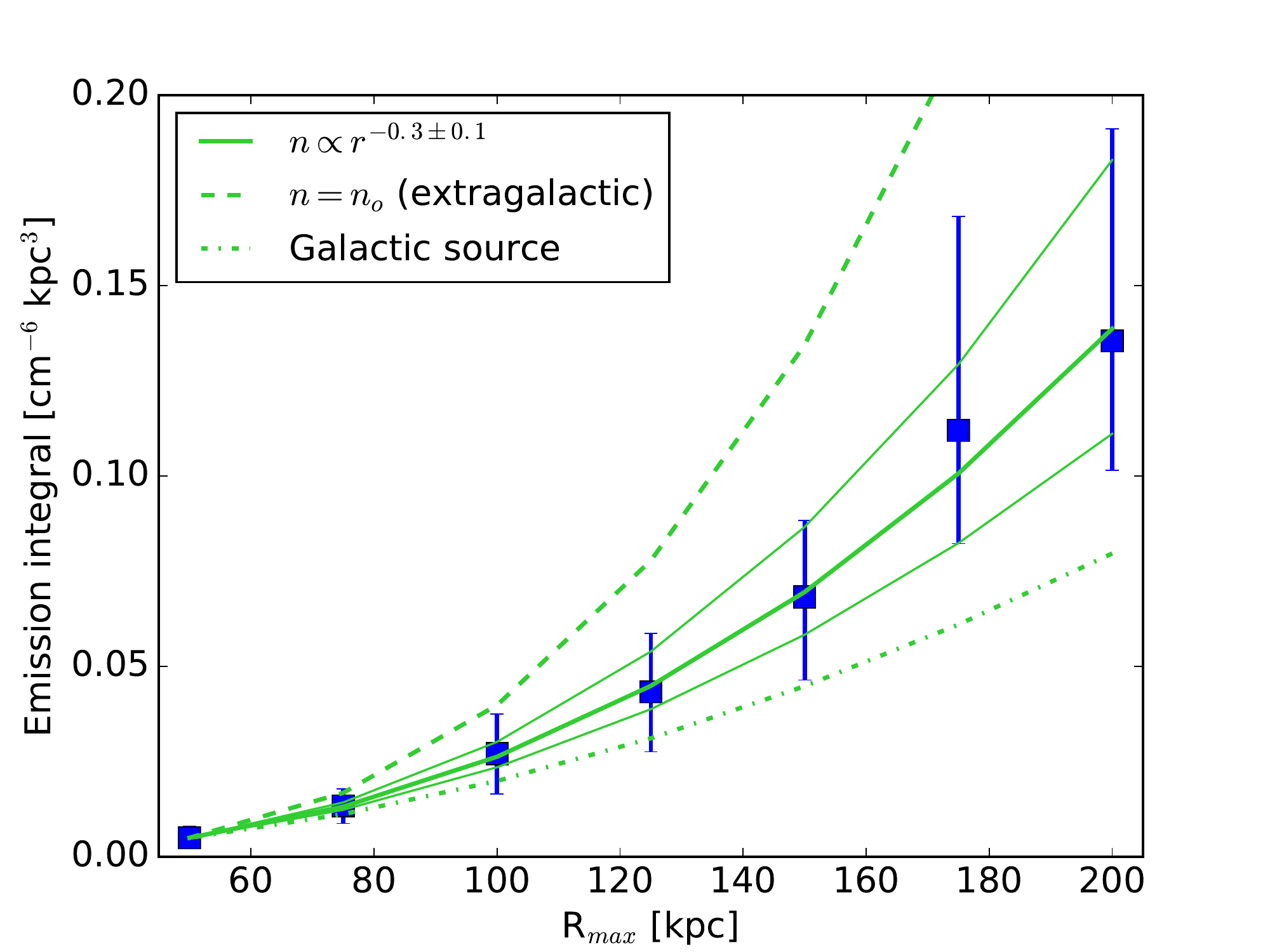}
    \caption{The emission integral as a function of enclosed volume. The profile for any Galactic source, extragalactic source unrelated to NGC\,3221, and the CGM of NGC\,3221 have been shown in green lines. The first two possibilities are ruled out by the data, physically confirming the presence of the CGM.}
    \label{fig:EI}
\end{figure}

The emission integral (EI = $\int \textcolor{black}{n_en_H} dV$ = $4\pi 10^{14} D^2 \times$ norm; where D is the comoving distance to the target) can be used to distinguish between the CGM of NGC\,3221 and any other source of confusion. A local Galactic diffuse source would have uniform surface brightness over the field. In that case, EI will be proportional to the projected area: $\int \textcolor{black}{n_en_H} dV$ = ($\int \textcolor{black}{n_en_H} dl) \times A$. For any extragalactic source unrelated to NGC\,3221, the density will be independent of the projected distance from NGC\,3221. EI will be proportional to the projected volume in this case: $\int \textcolor{black}{n_en_H} dV$ = $\textcolor{black}{n_en_H} \times V$. We find that the EI measurements do not agree with either of these scenarios (Figure \ref{fig:EI}). A simple power-law density profile: $n = n_o \Big(\frac{r}{r_o}\Big)^{-\alpha}$ with $\alpha = 0.3\pm0.1$ is consistent with the data. The exact shape of the density profile is not relevant here, nor do we argue that the shape is a power-law. A declining density profile from the galaxy center  shows that the diffuse medium is spatially correlated with NGC\,3221, thus physically confirming the signal to come from the CGM of NGC\,3221.  

\subsection{Comparison with earlier analysis}
\textcolor{black}{\noindent Below, we discuss our result in the context of previous findings from the same \xmm~data and our \suzaku~data. 
\subsubsection{Comparison with \suzaku}}
\noindent Excited by the confirmed significant detection of the CGM of NGC\,3221, we now perform the spectral analysis in differential annuli. First, we compare our results with those from \suzaku~\citep{Das2019b}. 

\begin{figure}[h]
    \centering
    \includegraphics[scale=0.425]{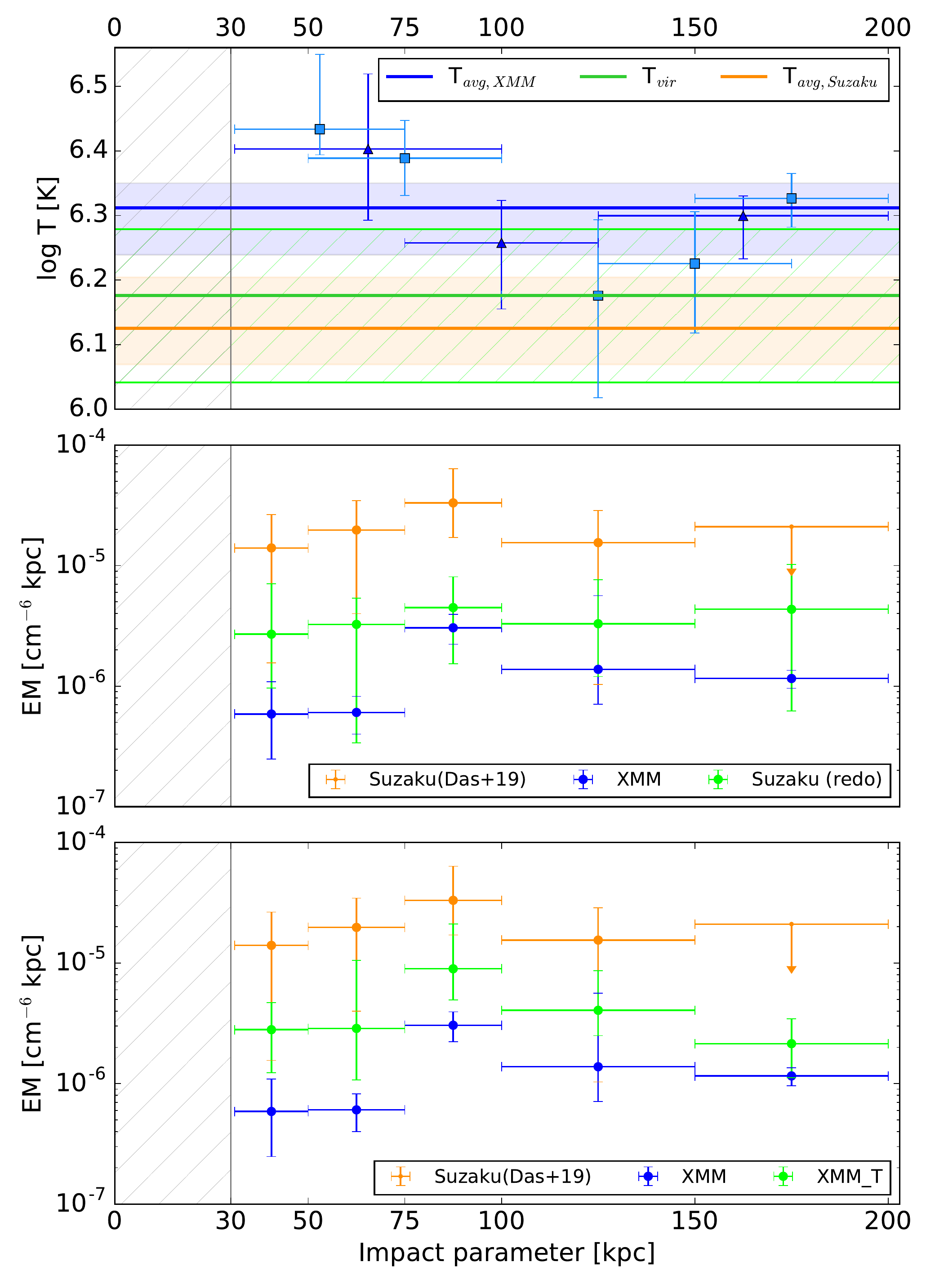}
    \caption{The temperature profile (top) and emission measure profile (bottom 2) of the warm-hot CGM of NGC\,3221 obtained from the \suzaku~\citep{Das2019b} and \xmm~data (this work). The hatched area in the left denotes to the disk and the extraplanar region, excluded from the analysis here. See text for details.}
    \label{fig:EM}
\end{figure}

\noindent In Figure \ref{fig:EM}, we have plotted temperature (top panel) and EM (second and third panels) as a function of the impact parameter, together with the measurements in  \cite{Das2019b}.  We derive the emission measure (EM) from the normalization parameter of the CGM spectrum ($EM = \int \textcolor{black}{n_en_H} dl = \frac{4\pi}{\Omega} \times 10^{14} \times norm$; where $\Omega$ is the solid angle projected by the annulus). 

The emission measure values from \xmm~beyond 100\,kpc are consistent with the measurement from \suzaku~within error, including the upper limit. However, the EM within 100\,kpc from \xmm~and \suzaku~differ by almost an order of magnitude. As the CGM emission signal depends sensitively on the XRB, the difference in the values of EM is likely due to the difference in the XRB. \cite{Das2019b} had used two off-fields $\sim$2$^\circ$ away from the galaxy-field to estimate the XRB. Modeling the XRB from the same FoV, as we do here, would minimize the spatial variation in the XRB (most likely arising from the MWH component). To test whether the differences in EM are indeed due to differences in XRB, we re-fit the \suzaku~data with the parameters of LHB and MWH frozen at their best-fitted values obtained from the \xmm~data. We also froze the temperature of the CGM of NGC\,3221 to the best-fit \xmm~values and re-derived the emission measures (labeled as `Suzaku (redo)' in the second panel of Figure \ref{fig:EM}). The \suzaku~EM values are now closer to our measurements from \xmm, and consistent within error. This shows that the same spectral model can fit both the \xmm~and \suzaku~data; this confirms that the major difference in EM in this work and \cite{Das2019b} is due to the different XRBs used to model the spectra.

The temperature obtained from \xmm~and \suzaku~are also significantly different within 100kpc (Figure \ref{fig:EM}, top panel). In our spectral modeling, temperature and emission measure are anticorrelated at a given metallicity. It is therefore possible that the temperature difference contributes to the differences in EM between \xmm~ and \suzaku~ data. To test this, we refit the \xmm~data by freezing the temperature of the CGM of NGC\,3221 at its best-fitted value obtained from \suzaku. The emission measures (labeled as `XMM$\_$T' in the third panel of Figure \ref{fig:EM}) are now consistent with those obtained by \cite{Das2019b}. This shows that the difference between the EM values in this work and \cite{Das2019b} can be explained by the difference in the XRB and the temperature of the CGM.  

While the revised EM values from \suzaku~are consistent with those from \xmm~within error, there are still residual differences, with the best-fit EM values systematically larger in \suzaku. This is likely due to following two factors:\\
1) Temporal variation in the foreground: The observed EM differences are possible if the XRB during the \xmm~observation was smaller than that during the \suzaku~observation. Indeed, the solar activity and the proton flux from solar wind during \xmm~observation were lower than those during \suzaku~data \citep{Sekiya2014}. The foreground during the \suzaku~observation might therefore be higher. \textcolor{black}{However, the solar wind charge exchange (SWCX) is not resolvable from LHB at the spectral resolution of \suzaku/XIS \citep{Yoshino2009} and it is challenging to model SWCX along with other foreground components in a shallow data \citep{Henley2015b}. Due to the lack of ``\textit{blank sky}" in the galaxy-field of \suzaku, quantifying the exact contribution of SWCX in the XRB of \suzaku~data is not possible. This shows how crucial it is to extract the XRB from the same observation as the target to alleviate the indefinite temporal variation of the XRB. The \xmm~ results are not affected by the temporal variation of the XRB; these are reliable values}.   \\
\textcolor{black}{2) Difference in CXB and spilled photons: The emission from unresolved extragalactic sources fainter than the detection limit of the telescope is modeled as CXB. The sensitivity of \suzaku~is lower than that of \xmm. Therefore, the flux down to which point sources are removed is higher in \suzaku~than that in \xmm. So the CXB of \suzaku~and \xmm~are unlikely the same. Also, the size of the point source regions were same as the PSF of \suzaku/XIS \citep[half power diameter $\approx2'$,][]{Das2019b}, i.e., only $\approx60\%$ of the flux from the point sources could be removed. However, in the \xmm~data we set the parameters of point source detection such that we can remove $\approx 83\%$ flux from the point sources (see \S\ref{sec:reduction}). This makes the contamination by spilled photons outside the removed regions in \xmm~much smaller than that in \suzaku. The excess emission measure of the CGM measured with \suzaku~compared to that in \xmm~can therefore be, at least partially, attributed to the different CXB and different amount of spilled photons from removed point sources.}\\
3) SPB variation over the the detector plane: We have assumed that the SPB is uniform over the detector plane, but the ESAS manual suggests that SPB may increase toward the edge of the detector. As we have assumed a uniform SPB obtained by modeling the ``\textit{blank sky}", we might overestimate the XRB and underestimate the CGM of NGC\,3221 at small radii toward the center of the detector plane. 

The average temperature derived from the \xmm~data is shown by the blue line in the top panel of Figure \ref{fig:EM}, and that from the \suzaku~data is shown by the orange line. The two differ by $2\sigma$, again because of the differences in the XRB. However, both are consistent with the virial temperature, modulo the huge uncertainty in the latter (green line in the top panel of Figure \ref{fig:EM}).

Thus the differences in the measured quantities from \xmm~and \suzaku~data can be understood by a combination of spatial and temporal differences in the X-ray background (XRB). As the CGM signal accounts for only $\approx$10\% of the data, it is very sensitive to the shape and the amplitude of the XRB. 
\textcolor{black}{
\subsubsection{Comparison with \xmm}
\noindent \cite{Tullmann2006} observed  NGC\,3221 with the same \xmm~data. Since they did not detect the CGM of this galaxy while we have, it is imperative to understand  why. In  their  imaging  analysis, \citet{Tullmann2006}  performed  contour  mapping  of  the  soft X-ray  image  of  the  galaxy,  including  the  disk. The surface  brightness  of  the  disk  of NGC\,3221 and  the  foreground  (local/Galactic) sources is much higher than that of the halo gas, therefore the contours were not sensitive to the faint extended emission.  In our analysis, we removed the inner 30\,kpc region around the galaxy (\S\ref{sec:redans}), and focused on the region beyond that.  This ensured that the CGM X-ray emission we study is not contaminated by the galaxy itself.  We have performed a detailed and intricate spectral analysis, where the ``\textit{on-source}" is simultaneously fitted with the ``blank-sky" to estimate the X-ray background (figure \ref{fig:spec}). This was crucial in extracting the weak CGM signal, but there is no spectral analysis in \citet{Tullmann2006}.  To  conclude,  we have detected  the  CGM  of  NGC\,3221,  but \citet{Tullmann2006} did not, because of a combination of differences in data extraction and spectral analysis.}

\subsection{Temperature and Emission Measure Profiles}\label{sec:modeling}

\noindent In the top panel of Figure \ref{fig:EM}, we have also plotted temperatures derived from large annuli (shown in blue triangles); these are within $1\sigma$ of the volume averaged temperature.  However, the temperatures obtained from smaller annuli (shown in blue squares) are inconsistent with a constant value. We find a clear trend of temperature decreasing by half a dex from 30 to 125\,kpc, and slowly increasing by a factor of $\sim$2 from 125 to 200\,kpc. Thus the warm-hot CGM of NGC\,3221 is approximately isothermal over the entire volume, but is not isothermal in finer details. The average temperature is consistent with the virial temperature of NGC\,3221, but the temperature within 100\,kpc is indicative of a super-virial temperature. Therefore, while the CGM on average is in thermal and hydrostatic equilibrium, the inner halo might deviate from that. This is an exciting new discovery from the \xmm~ data. Such a super-virial temperature has been observed in the CGM of the Milky Way both in emission and absorption \citep{Nakashima2018,Das2019a,Das2019c}, but this is the first time we are observing it in the CGM of any external spiral galaxy. This is also the first time a temperature gradient has been observed in the CGM of any spiral galaxy. Such a gradient and deviation from virial conditions have been predicted in semi-analytic and numerical models \citep{Maller2004,Pezzulli2017}, but never observed before.
Given the large errors in temperature and the wide spatial bins, we do not attempt to fit a profile to the temperature gradient; much higher quality data is required for that. In \cite{Das2019b} on \suzaku~data, the analysis was done only with ``cumulative"  annuli, and no such trend in temperature could be detected. Any temperature variation is washed out in large annuli in both \xmm~and \suzaku~data.  

\begin{figure}[h]
    \centering
    \includegraphics[trim = 10 0 0 40,clip,scale=0.45]{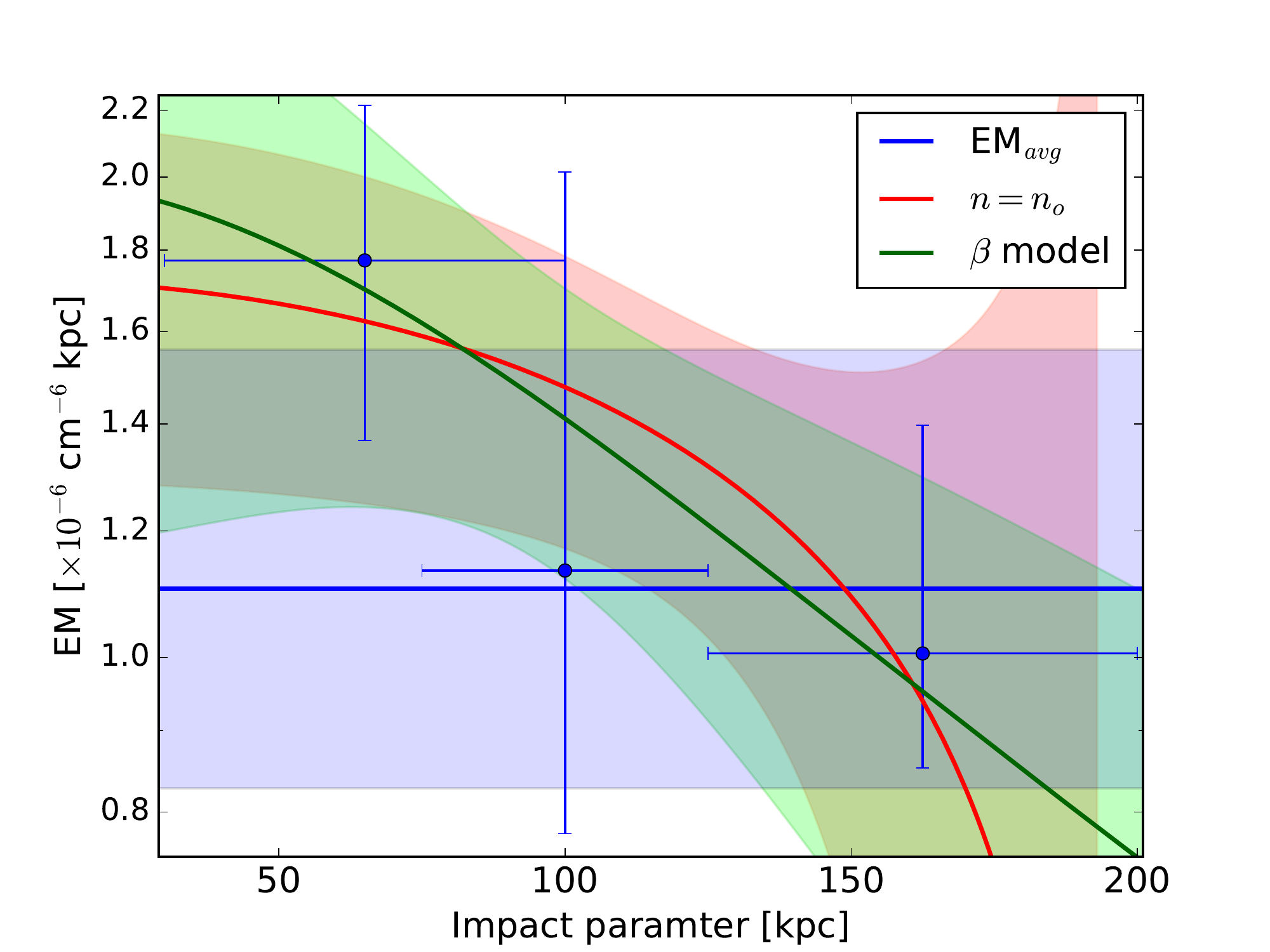}
    \caption{The emission measure profile of the warm-hot CGM of NGC\,3221. Additionally, we plot the volume averaged emission measure of the CGM. See text for details.}
    \label{fig:model}
\end{figure}
For emission measure, we consider only those values which are well-constrained; here we see a clear trend with impact parameter even in large spatial bins (Figure \ref{fig:model}), although they are consistent with the volume average of the EM within 1$\sigma$.  Once again, to check for the consistency with the \suzaku\ data, 
we fit the emission measure profile with two models: \citep[see their \S4.3]{Das2019b}.\\
A) A truncated constant-density model:  The best-fitted values are -- density $n_o$ = 6.7$\pm$1.2$\times$10$^{-5}$ cm$^{-3}$, spatial extent R$_{out}$ = 194$\pm$42\,kpc (Figure \ref{fig:model}, labeled as $n=n_o$). This value of R$_{out}$ is consistent with R$_{out}=175\pm2$ in \citet{Das2019b} and is of the order of the virial radius $R_{vir}=245^{+51}_{-77}$ kpc \citep[Hyperleda catalog,][]{Makarov2014}. Note that the constant density profile is consistent with the power-law profile of index $\alpha = 0.3\pm0.1$ (\S\ref{sec:confirm}) within $3\sigma$. \\
B) A $\beta$ model \citep{Sarazin1986}: 

\begin{equation}
    n = n_o \times \Big( 1 + (r/r_c)^2 \Big)^{-3\beta/2}
\end{equation}
We fit the model by freezing $\beta$ at $0.5$ (Allowing $\beta$ to vary did not produce any significant change in \cite{Das2019b}, so we do not try that here). The best-fitted values are -- central emission measure $EM_o$ = 2.0$\pm$0.4$\times$10$^{-6}$ cm$^{-6}$kpc, core radius r$_{c}$ = 155$\pm$46\,kpc (Figure \ref{fig:model}, labeled as $\beta$ model). The core radius is consistent with that obtained by \cite{Das2019b}. Also, $r_c$ is consistent with $R_{out}$ within uncertainties. That implies the density is practically flat out to a large radius, consistent with the constant density (or a power-law profile with small index). 

Thus we can fit the observed EM profile with either a $\beta$-model, or constant density model, but we cannot determine which model better represents the data. Deeper data would allow us to obtain both the temperature and the EM at finer spatial resolution and with higher precision, thus enabling us to prefer one density model over the other. 
\subsection{\textcolor{black}{Baryon content}}
\noindent \textcolor{black}{In order to determine the baryon census of NGC\,3221, we  consider  the  warm-hot  CGM  mass (modulo  metallicity), the mass of other phases of CGM, the mass of stellar and ISM components, and the virial mass.}

The mass of the constant density model at solar metallicity is $M = 5.7^{+6.5}_{-3.5} \times 10^{10} \Big(\frac{Z}{Z_\odot}\Big)^{-0.5}$ M$_\odot$. At solar metallicity, the central electron density of the $\beta$-model is $n_e \approx 10^{-4} cm^{-3}$ and the mass is $M = 5.3^{+1.0}_{-1.2} \times 10^{10} \Big(\frac{Z}{Z_\odot}\Big)^{-0.5}$ M$_\odot$. The masses from two models are similar, showing that the mass measurement is convergent. 

At $\frac{1}{3}$Z$_\odot$ metallicity \citep[the median metallicity of cool CGM at low redshift,][]{Prochaska2017}, the mass escalates to M$_{hot,halo}$ = $10\pm2\times10^{10}$ M$_\odot$. This is comparable with the stellar mass of NGC\,3221 : M$_\star$ = 10.0$\pm$1.3 $\times$10$^{10} M_\odot$ \citep{Lehmer2010}. \textcolor{black}{By comparing M$_{hot,halo}$ with other baryon components in the disk and the halo of NGC\,3221 (see Table 4 of \cite{Das2019b} and references therein), we find that it is one of the most massive baryon components of the galaxy. This is consistent with the findings of \cite{Gupta2012,Nicastro2016b} in the case of the Milky Way. By adding all baryon components, we find M$_{b,tot}$ = $34\pm5\times10^{10}$ M$_\odot$.}

\textcolor{black}{We estimate the virial mass M$_{vir}$ of NGC\,3221 from the maximum rotational velocity of the gas around the galaxy \citep[Hyperleda catalog,][]{Makarov2014} using baryonic Tully-Fisher relation for cold dark matter cosmology \citep{Navarro1997}, and obtain M$_{vir}$ =  3.12$\pm$1.40$^{+1.34}_{-2.57}$ $\times$10$^{12}$  M$_\odot$.
}

\textcolor{black}{The total baryon fraction is $f_b = \frac{M_{b,tot}}{M_{vir}} = 0.108\pm0.051$(statistical) $^{+0.091}_{-0.049}$(systematic). This is consistent with the cosmological baryon fraction $f_{cosmo} = 0.157\pm0.001$ \citep{Planck2016}. Thus, the warm-hot CGM clearly closes the baryon budget of NGC\,3221. }

\textcolor{black}{The uncertainty in the baryon fraction is three-fold.  First, the statistical uncertainty of the detected signal is propagated to the density profile and hence, the mass
of the warm-hot CGM. Second, the metallicity adds
a factor of few systematic uncertainty to the mass of
the warm-hot CGM. The uncertainty in the virial mass,
however, is the major source of uncertainty in $f_b$.}

\section{Summary, Conclusion and future directions}
\noindent In this paper, we have outlined an efficient, rigorous and well-defined method to extract the faint signal from the CGM of a galaxy using the archival \xmm~data of NGC\,3221. Following this method, we 1) determine the confidence-based spatial extent of the emission signal and the overall detection significance, 2) physically confirm the signal as coming from the CGM od NGC\,3221, and 3) obtain the temperature and the surface brightness profiles, even from  shallow data. Below, we summarize our science findings:
\begin{itemize}
    \item We have detected the warm-hot CGM of NGC\,3221 from the 30--200\,kpc region at $> 4\sigma$ significance. This confirms the 3.4$\sigma$ detection of the same signal by \cite{Das2019b} using \suzaku~data. 
    \item The signal is detected with $>99.6\%$ confidence beyond 150\,kpc. This implies that the CGM is truly extended to a significant fraction of the virial radius of NGC\,3221.
    \item There is a clear temperature gradient out to 100--125\,kpc, showing that the entire CGM is not isothermal. This is an exciting new result from this study. Additionally, we find that the temperature in the inner halo is super-virial.
    \item The EM profile is not well-constrained, but is consistent with a $\beta$ model or with a flat density profile.
    \item The warm-hot CGM is massive: $10\pm2 \times 10^{10}\Big(\frac{Z}{0.3Z_\odot}\Big)^{-0.5}$ M$_\odot$. It is one of the most massive baryon components of NGC\,3221.
\end{itemize} 

Thus, we confirm the discovery of the massive extended warm-hot CGM around the first external L$^\star$ galaxy. It can account for all of its galactic baryons, modulo the huge uncertainty in the virial mass. Deeper data with higher S/N and finer radial bins is required to provide a well-constrained temperature and density profile, and so understand the thermal and dynamical history of the CGM.

It is essential to study the CGM of galaxies with a broad range of M$_\star$, SFR and M$_{vir}$ to understand the key parameter governing the CGM. At present, \textit{XMM-Newton} and \textit{eROSITA} are the most suitable missions to detect the faint emission from the warm-hot CGM because of large effective area and large FoV. On a longer timescale, upcoming missions (e.g. \textit{XRISM, Athena}) in the next decade and beyond will offer an outstanding opportunity to observe the warm-hot diffuse medium in unprecedented detail. This will bring us closer to understanding the galaxy formation and evolution.
\section*{Acknowledgement}
\label{sec:conclude}
We gratefully acknowledge the NASA grant NNX16AF49G to SM. AG acknowledges support through the NASA ADAP grant 80NSSC18K0419.  

%

\vspace{5mm}
\facilities{\xmm}


\software{SAS v17.0.0 \citep{Snowden2004}, HeaSoft v6.17 \citep{Drake2005}, NumPy v1.11.0 \citep{Dubois1996}, Scipy v0.17.0 \citep{Oliphant2007}, Matplotlib v1.5.3 \citep{Hunter2007}, DS9 \citep{Joye2003}}





\bibliographystyle{aasjournal}



\end{document}